\documentclass[12pt,a4paper]{article}
\usepackage{amsmath,amssymb,url,cite,ifpdf,slashed,multirow,tabularx}

\ifpdf                                     
  \usepackage{graphicx}                    
\else                                      
  \usepackage[dvipdfmx]{graphicx}          
\fi

\usepackage[colorlinks=true, linkcolor=blue, citecolor=blue,urlcolor=blue]{hyperref} 

\def\thefootnote{\ifnum\c@footnote>\z@\textasteriskcentered\@arabic\c@footnote\fi}
\makeatletter
\renewcommand{\footnoterule}{%
\kern-3\p@
\hrule width 0.4\columnwidth
\kern 2.6\p@}
\def\thefootnote{\ifnum\c@footnote>\z@\@arabic\c@footnote\fi}
\makeatother

\allowdisplaybreaks

\newcommand{\TeV}{\,{\rm TeV}}
\newcommand{\GeV}{\,{\rm GeV}}

\newcommand{\invfb}{\,{\rm fb^{-1}}}

\def\be{\begin{equation}}
\def\ee{\end{equation}}
\def\beq{\begin{eqnarray}}
\def\eeq{\end{eqnarray}}

\def\({\left(}
\def\){\right)}
\def\<{\langle}
\def\>{\rangle}

\makeatletter
\newcommand{\@authornote}[2]{{\def\thefootnote{\fnsymbol{footnote}}\setcounter{footnote}{#1}#2\setcounter{footnote}{0}}}
\newcommand{\authornotemark}[1]{\@authornote#1{\addtocounter{footnote}{-1}\footnotemark}}
\newcommand{\authornotetext}[2]{\@authornote#1{\footnotetext{#2}}}
\makeatother

\begin{document}

\begin{titlepage}

\begin{flushright}
UT--14--01\\
January, 2014
\end{flushright}

\vskip 1.5 cm
\begin{center}

{\Large \bf
Future Prospects for Stau   in Higgs Coupling  \\[0.4em] to Di-photon}

\vskip .95in

{\large
\textbf{Motoi Endo},
\textbf{Teppei Kitahara}, and
\textbf{Takahiro Yoshinaga}
}
\vskip 0.4in

{\large
{\it 
Department of Physics,  Faculty of Science, \\[0.4em]
University of Tokyo, Bunkyo-ku, Tokyo 113-0033, Japan
}}

\end{center}
\vskip .95in

\begin{abstract}

We study future prospects of the stau which contributes to the Higgs coupling to di-photon. 
The coupling is sensitive to new physics and planned to be measured at percent levels in future colliders. 
We show that, if the excess of the coupling is measured to be larger  than $4\%$, the lightest stau is predicted to be lighter than about $200\GeV$  by taking  vacuum meta-stability conditions into account.
Such a stau can be discovered at ILC.
Moreover, we show  how accurately the stau contribution to the coupling  can be reconstructed from the information that is available at ILC. 
We also argue that, if the stau mixing angle is measured, the mass of the heaviest stau can be predicted by measuring the Higgs coupling, even when the heaviest stau is not yet discovered at the early stage of ILC.

\end{abstract}

\end{titlepage}
\setcounter{page}{1}
\hrule
\tableofcontents
\vskip .2in
\hrule
\vskip .4in

\section{Introduction} \label{sec:introduction}

The electroweak oblique corrections, which are self-energies of the electroweak vector bosons, are sensitive to new physics.
Similarly, loop-induced Higgs couplings, i.e., the Higgs boson coupling to di-photon, di-gluon or $Z\gamma$, constrain the new physics, and they are called the Higgs oblique corrections \cite{Gori:2013mia}.
In the Standard Model (SM), these couplings are prevented by the gauge symmetry at the tree level and induced at  radiative levels. 
Therefore, the new physics may be probed indirectly by measuring the loop-induced Higgs couplings in future.

Particularly, the Higgs coupling to di-photon is important. 
In the SM, it is dominated by one-loop contributions of the electroweak vector bosons and the top quark. 
If new physics contains charged particles that couple to the Higgs boson, they contribute to the Higgs coupling to di-photon at radiative levels. 
Hence, the Higgs coupling is sensitive to the new physics contributions. 
If such new particles exist, it is expected that deviations from the SM prediction are observed.

In this letter, we parametrize the deviation of the $126\GeV$ Higgs boson coupling from the SM prediction as
\begin{equation}
\kappa_{A} = \frac{g_{hAA}}{g_{hAA}(\textrm{SM})} = 1 + \delta \kappa_{A},
\end{equation}
where $g_{hAA}$ is the Higgs coupling to the $A\bar{A}$ particles, and the new physics contribution is represented by $\delta \kappa_{A}$.
At present, the Higgs coupling to di-photon, $\kappa_{\gamma}$, has been measured with the uncertainty of 15\% ($1\sigma$) at ATLAS \cite{Aad:2013wqa} and 25\% at CMS \cite{CMS:yva}. 
The results are consistent with the SM prediction, though they are not yet precise enough to probe new particle contributions. 
In future, LHC will accumulate the luminosity $\mathcal{L} \sim 300\invfb$ at $\sqrt{s}=14\TeV$, and further upgrade is proposed for $\sim 3000\invfb$ at High-Luminosity LHC (HL-LHC). 
The accuracies of $\kappa _{\gamma}$ are, then, expected to be about 7\% and 5\% at $300\invfb$ and $3000\invfb$, respectively \cite{ATLAS:2013hta}. 
Since the errors are dominated by systematic uncertainties, the accuracies could be improved by reducing them. 
It is recently argued that the sensitivity can be improved well, once the international $e^+e^-$ linear colliders (ILC) will be constructed \cite{Peskin:2013xra}. 
At LHC, the ratio of the branching fractions of $h \to \gamma\gamma$ and $h \to ZZ^*$ will be measured very precisely. 
At ILC, the Higgs couplings, including $\kappa_{Z}$ but $\kappa_{\gamma}$, can be measured at (sub) percent levels \cite{Asner:2013psa}. 
The joint analysis of HL-LHC and ILC enables us to realize the accuracy of $\kappa_{\gamma}$ of about 2\% \cite{Peskin:2013xra}. 
Here, it is assumed that the uncertainty of $\textrm{Br}(h \to \gamma \gamma)/\textrm{Br}( h \to ZZ^{\ast})$ is 3.6\% from HL-LHC, and ILC runs at $\sqrt{s}=250\GeV$ and $\mathcal{L} = 250\invfb$. 
The direct measurement of $\kappa_{\gamma}$ at ILC is not so precise that of LHC, 
 because the luminosity is limited. 
If more luminosity is accumulated, e.g., $\mathcal{L} = 2500\invfb$ at $\sqrt{s}=1\TeV$, the accuracy of the direct measurement of $\kappa_{\gamma}$ can become 1.9\% at ILC \cite{Peskin:2013xra, Asner:2013psa}, and
the accuracy of  the joint analysis of HL-LHC and ILC  can be better than 1\% \cite{Peskin:2013xra}. 
They are very precise, and it is expected that new charged particles could be probed  by measuring $\kappa_{\gamma}$.

In this letter, let us consider a situation that an excess of $\kappa_{\gamma}$ is measured in HL-LHC and ILC. 
Then, it is important to reveal which particle is responsible for the anomalous excess. 
A lot of models that affect $\kappa_{\gamma}$ have been proposed. 
Among them, a scalar partner of the tau lepton (stau) in supersymmetry (SUSY) models is one of the most motivated candidates. 
The Higgs coupling to di-photon is enhanced when the staus are light and when the mixing of left-handed and right-handed staus is large \cite{Carena:2011aa, Cao:2012fz, Carena:2012gp}. 
Since such staus are characteristic, they may be discovered and investigated in future colliders. 
In this letter, we study properties of staus that are responsible for the $\kappa_{\gamma}$ excess.
In particular, it will be shown that, by taking the vacuum meta-stability condition into account, the lightest stau is predicted to be discovered at ILC, if the deviation of $\kappa_{\gamma}$ is large enough to be detected.  
Therefore, we will discuss that the stau contribution to $\kappa_{\gamma}$ can be probed at ILC. 

Once the stau is discovered at ILC, its properties as well as the mass will be determined precisely \cite{Bechtle:2009em}.
It may be possible to investigate whether the properties are consistent with the contribution to $\kappa_{\gamma}$. 
If the heaviest stau as well as the lightest one is discovered, the stau contribution can be reconstructed directly by using the information which is available from the measurements. 
We will show that the contribution can be reconstructed precisely at ILC. 
Since the uncertainty is comparable to or less than that of the measured $\kappa_{\gamma}$, it is possible to test whether the excess of $\kappa_{\gamma}$ originates in the stau contribution. 
In addition, 
we will discuss that the mass of the heaviest stau can be predicted by measuring  the excess of $\kappa_{\gamma}$ and the stau mixing angle, even if the heaviest stau is not yet discovered at the early stage of ILC. 
This prediction could be tested in the next stage of ILC. 

This letter is organized as follows. 
In Sec.~\ref{sec:stau}, we will briefly review the stau contribution to $\kappa_{\gamma}$ and the vacuum meta-stability condition.
In Sec.~\ref{sec:mass}, the stau mass regions to deviate $\kappa_{\gamma}$ will be studied. 
In Sec.~\ref{sec:prospect}, stau properties will be investigated. 
The last section is devoted to the conclusion.

\section{Stau Contributions}\label{sec:stau}

In this section, we briefly review the stau contribution to $\kappa_{\gamma}$ and the vacuum meta-stability condition.
The stau contribution becomes sizable when the stau is light and when the left-right mixing parameter of the left-handed and right-handed staus is large \cite{Carena:2011aa, Cao:2012fz, Carena:2012gp}. 
Since too large left-right mixing parameter spoils the stability of our ordinary vacuum, the parameter is limited \cite{Casas:1995pd}. 
Thus, the stau contribution to $\kappa_{\gamma}$ is constrained by the vacuum meta-stability condition \cite{Carena:2012gp,Sato:2012bf}. 

Let us first specify the framework. We consider the setup that only the staus and the Bino are light among the SUSY particles, while the other SUSY particles are heavy. The Bino is introduced as the lightest SUSY particle.
This avoids cosmological difficulties of stable heavy charged particles. Also, the setup is consistent with the recent LHC results. The Higgs boson mass of $126\GeV$ favors heavy scalar top quarks (stops). Absent signals in direct SUSY searches restrict colored SUSY particles to being heavier than $\sim 1\TeV$. In this letter, the stau contribution to $\kappa_{\gamma}$ and the stau properties will be studied. The above assumption is minimal for this purpose. Contributions from the other SUSY particles will be discussed later.

Staus are characterized by the mass eigenvalues and the left-right mixing angle as follows
\begin{equation}
  m_{\tilde{\tau}_1},~
  m_{\tilde{\tau }_2},~
  \theta _{\tilde{\tau}}.
  \label{eq:parameters}
\end{equation}
In addition, some of the stau couplings depend on $\tan\beta$, which is a ratio of the vacuum expectation values (VEVs) of the up-type and down-type Higgs fields.
These parameters are related to the SUSY model parameters through the mass matrix,
\begin{equation}
  M_{\tilde{\tau}}^2 = 
  \begin{pmatrix}
    m_{\tilde{\tau}{LL}}^2
    & 
    m_{\tilde{\tau}{LR}}^2
    \\
    m_{\tilde{\tau}{LR}}^2
    & 
    m_{\tilde{\tau}{RR}}^2
  \end{pmatrix},
  \label{eq:MassMatrix}
\end{equation}
where $m_{\tilde{\tau}{LL,RR}}^2 = \tilde m_{\tilde{\tau}{L,R}}^2 + m_{\tau}^2 + D_{\tilde{\tau}{L,R}}$ with soft SUSY-breaking parameters, $\tilde m_{\tilde{\tau}{L}}^2$ and $\tilde m_{\tilde{\tau}{R}}^2$, and D-terms, $D_{\tilde{\tau}} = m_Z^2 \cos 2\beta (I^3_{\tau} - Q_{\tau} \sin^2 \theta_W)$. The left-right mixing parameter is $m_{\tilde{\tau}{LR}}^2 = m_\tau (A_\tau - \mu_H \tan\beta)$, where $A_\tau$ and $\mu_H$ are the scalar tau trilinear coupling and the Higgsino mass parameter, respectively. The mass matrix is diagonalized as $U_{\tilde\tau} {\cal M}_{\tilde{\tau}}^2 U_{\tilde\tau}^\dagger = {\rm diag}(m_{\tilde{\tau}_1}^2, m_{\tilde{\tau }_2}^2)$ by the unitary matrix,
\begin{equation}
  U_{\tilde\tau} = 
  \begin{pmatrix}
  \cos\theta_{\tilde\tau} & \sin\theta_{\tilde\tau} \\
 -\sin\theta_{\tilde\tau} & \cos\theta_{\tilde\tau}
  \end{pmatrix}.
  \label{eq:UnitaryMatrix}
\end{equation}
Here, $m_{\tilde{\tau}_1} < m_{\tilde{\tau }_2}$ is chosen. 
It is found that $m_{\tilde{\tau}{LR}}^2$ satisfies a relation,
\begin{equation}
  m_{\tilde\tau{LR}}^2 = 
  \frac{1}{2} (m_{\tilde\tau_1}^2 - m_{\tilde\tau_2}^2) 
  \sin 2 \theta_{\tilde\tau}.
  \label{eq:MixingAngle}
\end{equation}
On the other hand, the Bino almost composes the lightest neutralino, whose mass is written as $m_{\tilde\chi^0_1}$. 
Although the neutralinos are composed of the Wino and the Higgsinos as well as the Bino, the Wino is supposed to be decoupled, and the Higgsinos are heavy in order to deviate $\kappa_{\gamma}$ sizably (see below). 
In this letter, CP-violating phases are neglected. 

The Higgs coupling to di-photon is composed of the contributions from the SM particles and the staus.
Theoretically, $\kappa_{\gamma}$ is represented as
\begin{align}
  \kappa_{\gamma} = 
  \frac{|\mathcal{M}_{\gamma\gamma}({\rm SM}) + \mathcal{M}_{\gamma\gamma}(\tilde\tau)|}{|\mathcal{M}_{\gamma\gamma}({\rm SM})|},
  \label{eq:kappa}
\end{align}
where $\mathcal{M}_{\gamma\gamma}$ is related to the Higgs decay rate as
\begin{align}
  \Gamma(h \to \gamma\gamma) = 
  \frac{\alpha^2m_h^3}{1024\pi^3} \left| \mathcal{M}_{\gamma\gamma} \right|^2.
\end{align}
The right-hand side in Eq.~\eqref{eq:kappa} is dominated by the one-loop contributions. The stau contribution is given by \cite{Shifman:1979eb}
\begin{align}
  \mathcal{M}_{\gamma\gamma}(\tilde\tau) =
  \sum_{i=1,2} \frac{g_{h\tilde\tau_i\tilde\tau_i}}{m_{\tilde\tau_i}^2} 
  A_0^h (x_{\tilde\tau_i}).
  \label{eq:MgammaStau}
\end{align}
where $x_i = 4 m_i^2/m_h^2$. The definition of the loop function $A_0^h(x)$ is given in Ref.~\cite{Shifman:1979eb}. In the decoupling limit of heavy Higgs bosons, the stau-Higgs couplings are approximated as
\begin{align}
  g_{h\tilde\tau_{1}\tilde\tau_{1},h\tilde\tau_{2}\tilde\tau_{2}} =
  \frac{1}{2}(\delta m_{\tilde{\tau}{LL}}^2+\delta m_{\tilde{\tau}{RR}}^2) 
  \pm 
  \frac{1}{2}(\delta m_{\tilde{\tau}{LL}}^2-\delta m_{\tilde{\tau}{RR}}^2) \cos 2\theta_{\tilde\tau} 
  \pm 
  \delta m_{\tilde{\tau}{LR}}^2\, \sin 2\theta_{\tilde\tau},
  \label{eq:StauCoupling}
\end{align}
where the coefficients are
\begin{align}
  \delta m_{\tilde{\tau}{LL,RR}}^2 = 
  \frac{2}{v} (m_\tau^2 + D_{\tilde{\tau}{L,R}}),~~~
  \delta m_{\tilde{\tau}{LR}}^2 = \frac{1}{v} m_{\tilde{\tau}{LR}}^2.
\end{align}
Here, $v$ is the SM Higgs VEV, $v \simeq 246\GeV$. 
It is noticed that the mass scale of $\delta m_{\tilde{\tau}{LL}}^2$ and $\delta m_{\tilde{\tau}{RR}}^2$ is set by the EW scale, whereas that of $\delta m_{\tilde{\tau}{LR}}^2$ is by the SUSY parameter.
On the other hand, the SM contribution is dominated by the one-loop contributions of the electroweak vector bosons and the top quark as \cite{Shifman:1979eb}
\begin{align}
  \mathcal{M}_{\gamma\gamma}({\rm SM}) =
  \frac{g_{hWW}}{m_W^2} A_1^h (x_W) + 
  \frac{2g_{htt}}{m_t} \frac{4}{3} A_{1/2}^h (x_t),
  \label{eq:MgammaSM}
\end{align}
where the coefficients are $g_{hWW}/m_W^2 = 2g_{htt}/m_t = 2/v$. The definitions of the loop functions, $A_1^h(x)$ and $A_{1/2}^h(x)$, are given in Ref.~\cite{Shifman:1979eb}.
From Eqs.~\eqref{eq:MgammaStau} and \eqref{eq:MgammaSM}, it is found that $\kappa_{\gamma}$ is deviated from the SM prediction sizably when $m_{\tilde{\tau}{LR}}^2$ is large. In fact, $\delta m_{\tilde{\tau}{LR}}^2$ is proportional to $m_{\tilde{\tau}{LR}}^2$, and $\sin 2\theta_{\tilde\tau}$ becomes sizable when $m_{\tilde{\tau}{LR}}^2$ is large, according to Eq.~\eqref{eq:MixingAngle}. 
It is also noticed that $\mathcal{M}_{\gamma\gamma}(\tilde\tau)$ is enhanced when $\tilde\tau_1$ is light. On the contrary, heavy $\tilde\tau_2$ is favored to enhance it, because the contribution of $\tilde\tau_2$ destructively interferes with that of $\tilde\tau_1$.
Also, once $m_{\tilde{\tau}{LR}}^2$ is given, the stau contribution is insensitive to $\tan\beta$. 

It is important that $m_{\tilde{\tau}{LR}}^2$ is limited by the meta-stability condition of the ordinary vacuum. 
As noticed in Eq.~\eqref{eq:StauCoupling}, large $m_{\tilde{\tau}{LR}}^2$ increases trilinear couplings of the stau-Higgs potential and eventually makes the ordinary vacuum unstable.
Thus, the mixing parameter is constrained. 
The fitting formula of the vacuum meta-stability condition is known as \cite{Kitahara:2013lfa, Endo:2013lva} 
\begin{align}
\left|m_{\tilde\tau{LR}}^2 \right| 
& \leq 
\eta  
\bigg[
1. 01  \times 10^2 \GeV \sqrt{\tilde m_{\tilde{\tau}L} \tilde m_{\tilde{\tau}R}} 
+ 1.01 \times 10^2 \GeV  (\tilde m_{\tilde{\tau}L} + 1.03\, \tilde m_{\tilde{\tau}R}) 
\notag \\
& -2.27 \times 
10^4\GeV^2
+ \frac{2.97 \times 10^6\GeV^3}{\tilde m_{\tilde{\tau}L} + \tilde m_{\tilde{\tau}R}} 
- 1.14 \times 10^8\GeV^4 
  \left( \frac{1}{\tilde m^2_{\tilde{\tau}L}} +  \frac{0.983}{\tilde m^2_{\tilde{\tau}R}} \right) 
\bigg]. 
\label{eq:VacuumStability}
\end{align}
Here, the Higgs potential is set to reproduce $m_h = 126\GeV$. A scale factor, $\eta\ (\simeq 1)$, is introduced to take account of a weak dependence on $\tan\beta$. This comes from Yukawa interactions in the quartic terms of the scalar potential. Numerical estimation of $\eta$ is found in Fig.~2 of Ref.~\cite{Kitahara:2013lfa}. For instance, $\eta \simeq 0.90$ for $\tan\beta = 20$. 
By combining Eqs.~\eqref{eq:kappa} and \eqref{eq:VacuumStability}, the stau properties, in particular the stau masses, are determined.

\section{Stau Mass Region} \label{sec:mass}

\begin{figure}[tb]
 \begin{center}
 \includegraphics[width=70mm]{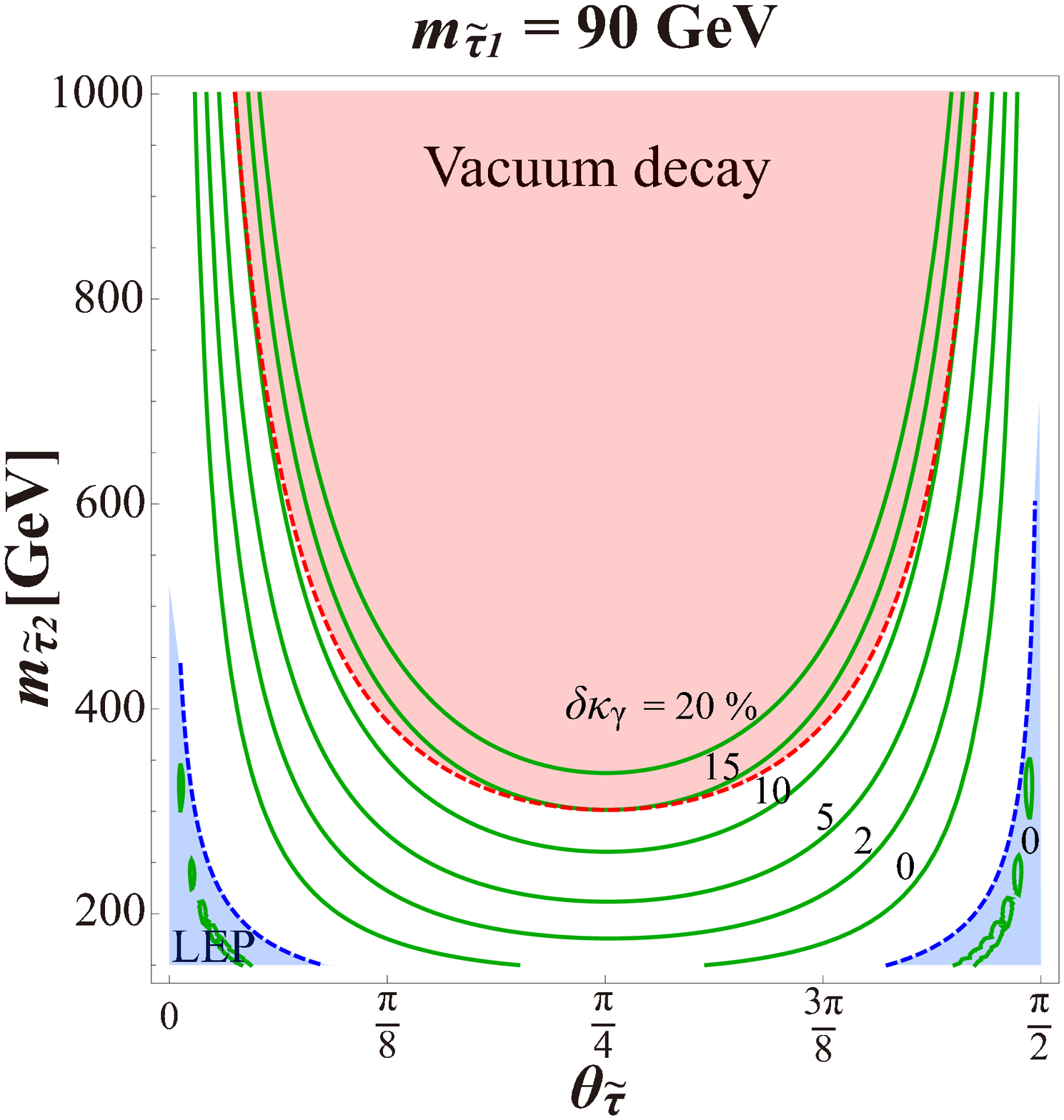} \hspace*{2mm}
 \includegraphics[width=70mm]{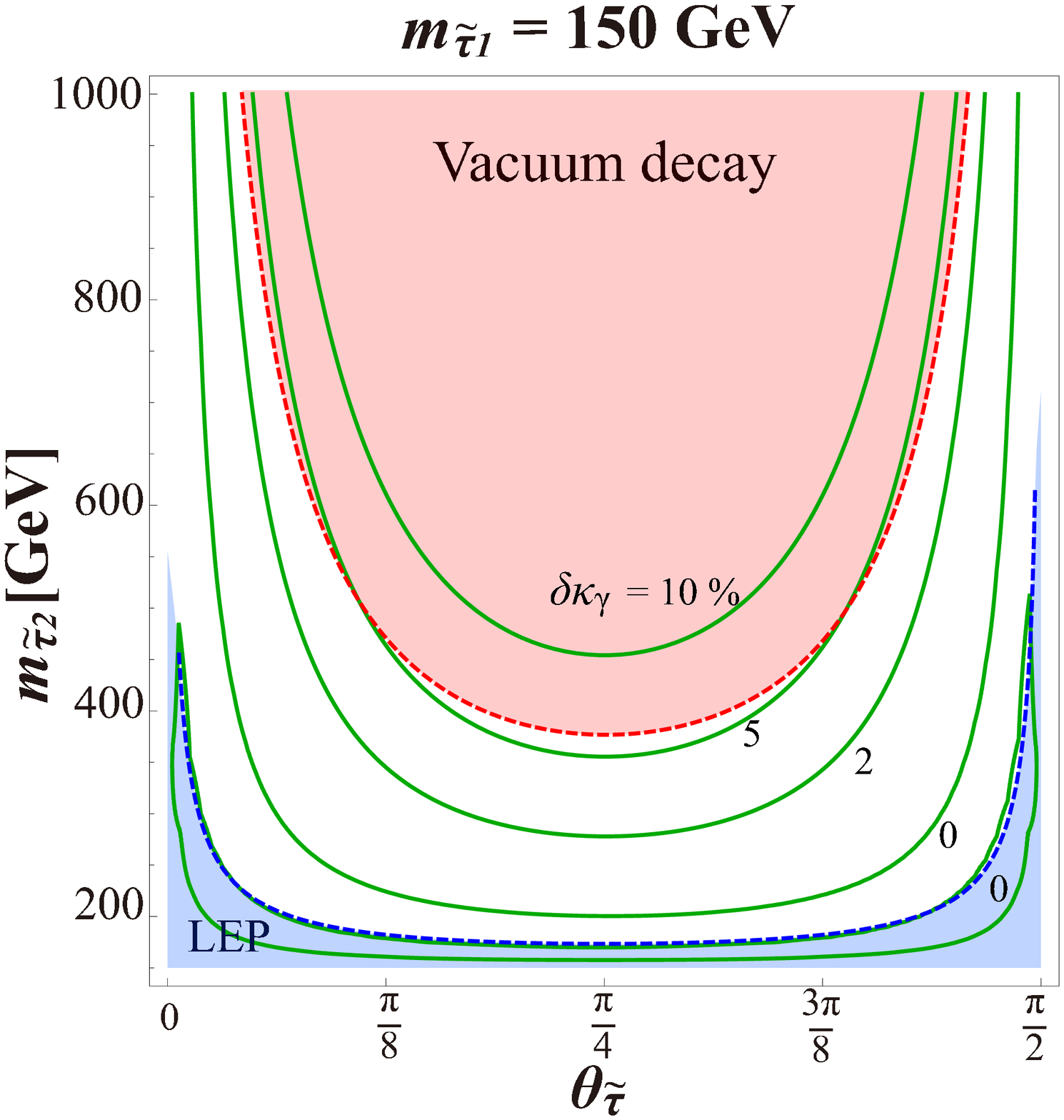} 
 \end{center}
 \caption{Contours of $\delta \kappa_{\gamma}$ are shown by the green solid lines. 
The lightest stau mass is $90\GeV$ (\textbf{left}) and $150\GeV$ (\textbf{right}). 
Here, $\tan\beta = 20$ and $A_\tau = 0$ are chosen. 
The red regions  are excluded by the vacuum meta-stability condition (\ref{eq:VacuumStability}). 
The blue regions are excluded by the chargino search at LEP.}
 \label{fig:mass_angle}
\end{figure}

In this section,
we study the stau mass region where the Higgs coupling  $\kappa_{\gamma}$ is deviated from SM prediction.
The stau contribution to the coupling is determined, once the stau parameters \eqref{eq:parameters} are given. 
They are constrained by the vacuum meta-stability condition. 

In Fig.~\ref{fig:mass_angle}, contours of $\delta \kappa_{\gamma} = \kappa_{\gamma} -1$ are shown by the green solid lines for given $m_{\tilde\tau_1}$ as a function of $\theta_{\tilde\tau}$ and $m_{\tilde\tau_2}$. 
The stau contribution depends on $\theta_{\tilde\tau}$ and is maximized when $\sin 2\theta_{\tilde\tau}$ is close to unity ($\theta_{\tilde\tau} \sim \pi/4$) for fixed $m_{\tilde\tau_1}$ and $m_{\tilde\tau_2}$. 
Also, $\delta \kappa_{\gamma}$ is enhanced by larger $m_{\tilde\tau_2}$.
On the other hand, if $\tilde\tau_2$ is very heavy, the stau contribution to $\kappa_{\gamma}$ becomes insensitive to $m_{\tilde\tau_2}$ and controlled by $m_{\tilde\tau_1}$ and $\theta_{\tilde\tau}$.

In Fig.~\ref{fig:mass_angle}, the red regions are excluded by the vacuum meta-stability condition. 
Eq.~\eqref{eq:VacuumStability} gives an upper bound on $m_{\tilde\tau{LR}}^2$ for given $m_{\tilde\tau_1}$ and $m_{\tilde\tau_2}$. 
Then, combined with Eq.~\eqref{eq:MixingAngle}, $\theta_{\tilde\tau}$ is constrained as a function of $m_{\tilde\tau_1}$ and $m_{\tilde\tau_2}$.
When $m_{\tilde\tau_2}$ is small, the angle is not limited by the vacuum meta-stability condition, and $\delta \kappa_{\gamma}$ is maximized when $\sin 2\theta_{\tilde\tau} = 1$ is satisfied. 
It is found that $\delta \kappa_{\gamma}$ becomes largest just below the red region with $\sin 2\theta_{\tilde\tau} = 1$ in each panel of Fig.~\ref{fig:mass_angle}.
On the other hand, the vacuum meta-stability condition constrains the stau mixing angle for large $m_{\tilde\tau_2}$.
The maximal value of $\delta \kappa_{\gamma}$ decreases, as $m_{\tilde\tau_2}$ increases.
When $\tilde\tau_2$ is very heavy, the vacuum meta-stability condition becomes insensitive to $m_{\tilde\tau_2}$ and determined by $m_{\tilde\tau_1}$. 
This is because, in the decoupling limit, $\tilde\tau_2$ does not contribute to the field configuration of the bounce solution to derive the vacuum meta-stability condition. 
Then, the maximal value of $\delta \kappa_{\gamma}$ is determined by $m_{\tilde\tau_1}$.

It is also noticed that the condition \eqref{eq:VacuumStability} is asymmetric under the exchange of the stau chirality, $\tilde\tau_L \leftrightarrow \tilde\tau_R$. 
However, the effect is negligibly small. 
In Fig.~\ref{fig:mass_angle}, it is found that the red region in $0 < \theta_{\tilde{\tau}} < \pi/4$ is almost coincide with that of $\pi/4 < \theta_{\tilde{\tau}} < \pi/2$.

In the analysis, $\tan\beta = 20$, $A_\tau = 0$ and $M_2 = 500\GeV$ are chosen, where $M_2$ is the Wino mass.
The stau contribution to $\kappa_{\gamma}$ and the vacuum meta-stability condition are almost independent of them, once $m_{\tilde{\tau}{LR}}^2$ is given. 
Rather, they are included in the definition of $m_{\tilde{\tau}{LR}}^2$ in association with the Higgsino mass parameter $\mu_H$. 
For fixed $m_{\tilde{\tau}{LR}}^2$, $\mu_H$ becomes smaller, as $\tan\beta$ increases. 
When the charginos are light, they can affect the Higgs coupling \cite{Carena:2011aa,Batell:2013bka}. 
Their contribution to $\kappa_{\gamma}$ is taken into account for completeness.
It is at most a few percents in the vicinity of the blue region and much less than 1\% around the red region in Fig.~\ref{fig:mass_angle}.
The blue region is already excluded by LEP \cite{CharginoLEP}, where the lightest chargino mass is less than $104\GeV$.

\begin{figure}[tb]
\begin{center}
 \includegraphics[width=120mm]{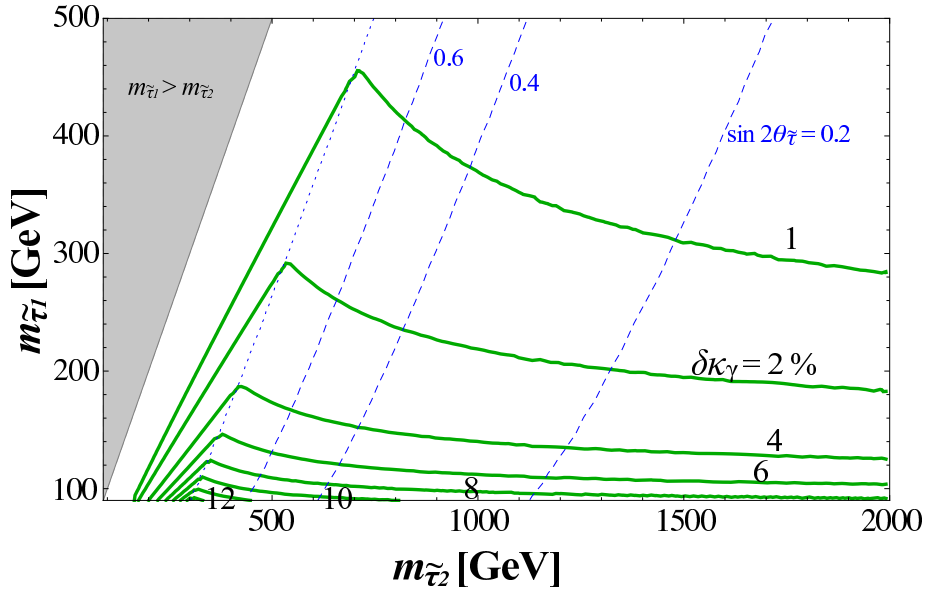}  
 \end{center}
 \caption{Contours of $\delta \kappa_{\gamma}$ are shown by the green solid lines.
At each point, $\delta \kappa_{\gamma}$ is maximized under the vacuum meta-stability condition.
Here, $\tan\beta = 20$, $A_\tau = 0$ and $\pi/4 \leq\theta_{\tilde{\tau}} < \pi/2$ are taken. 
The blue dashed lines are contours of $\sin 2 \theta_{\tilde{\tau}}$.
In the left region of the blue dotted line (the leftmost blue line), $\sin 2 \theta_{\tilde{\tau}} = 1$ is satisfied.}
 \label{fig:StauMass}
\end{figure}

Let us study the stau mass region. 
In Fig.~\ref{fig:StauMass}, contours of $\delta \kappa_{\gamma}$ are shown by the green solid lines. 
At each $(m_{\tilde\tau_1},m_{\tilde\tau_2})$, $\delta \kappa_{\gamma}$ is maximized with satisfying the vacuum meta-stability condition \eqref{eq:VacuumStability}. 
Each contour is composed of the two regions. 
In the left region of the peak, where $m_{\tilde\tau_2}$ is small, $\sin 2\theta_{\tilde\tau} = 1$ is satisfied. 
The stau contribution to $\kappa_{\gamma}$ is enhanced when $m_{\tilde\tau_2}$ is larger, as explained above.  
On the other hand, in the right region of the peak, $\theta_{\tilde\tau}$ is limited by the vacuum meta-stability condition. 
Here, $\sin 2\theta_{\tilde\tau}$ is less than unity. This is observed by the blue dashed lines, which are contours of $\sin 2\theta_{\tilde\tau}$ in Fig.~\ref{fig:StauMass}. 
As already found in Fig.~\ref{fig:mass_angle}, $\kappa_{\gamma}$ is enhanced, when $m_{\tilde\tau_2}$ is smaller. 

In the figure, $\tan\beta = 20$, $A_\tau = 0$ and $M_2 = 500\GeV$ are chosen. 
The results are almost independent of them except for the region in the vicinity of $m_{\tilde\tau_1} = m_{\tilde\tau_2}$.
When $m_{\tilde\tau_1}$ is very close to $m_{\tilde\tau_2}$, the Higgsinos become light, because the stau left-right mixing parameter tends to be small (see Eq.\eqref{eq:MixingAngle}).
Then, the charginos can contribute to $\kappa_{\gamma}$. 
Otherwise, their contribution is negligible in Fig.~\ref{fig:StauMass}. 
In the figure, it is also supposed that the lightest stau is mainly composed of the right-handed component, $\pi/4 \leq\theta_{\tilde{\tau}} < \pi/2$. 
As mentioned above, the stau mass region in Fig.~\ref{fig:StauMass} is almost insensitive to this choice. 

Currently, the measured values of $\kappa_{\gamma}$ at LHC are consistent with the SM prediction. 
The uncertainties are 15\% (ATLAS) \cite{Aad:2013wqa} and 25\% (CMS) \cite{CMS:yva}. 
As found in Fig.~\ref{fig:StauMass}, they are not precise enough to probe the stau contribution for $m_{\tilde\tau_1} > 100\GeV$.
In future, the sensitivity will be improved very well, as mentioned in Sec.~\ref{sec:introduction}.
It is expected that LHC measures $\kappa_{\gamma}$ at about 7\% and 5\% for the luminosities, $300\invfb$ and $3000\invfb$, respectively with $\sqrt{s}=14\TeV$ \cite{ATLAS:2013hta}. 
If the measurement of $\textrm{Br}(h \to \gamma \gamma)/\textrm{Br}(h \to ZZ^{\ast})$ at HL-LHC is combined with the measurements of the Higgs couplings at ILC, it was argued that the uncertainty of $\kappa_{\gamma}$ can be reduced to be about 2\% ($1\sigma$) at $250\GeV$ ILC with $\mathcal{L} = 250\invfb$ \cite{Peskin:2013xra}. 
If the luminosity is accumulated up to $2500\invfb$ at $1\TeV$ ILC, it has been estimated that the accuracy of $\kappa_{\gamma}$ can be better than 1\% \cite{Peskin:2013xra}. 

It is noteworthy that, once an excess of $\kappa_{\gamma}$ is measured, the mass region of staus are determined from Fig.~\ref{fig:StauMass}.
From the joint analysis of $250\GeV$ ILC and HL-LHC, $\delta \kappa_{\gamma}$ is expected to be measured with the uncertainty of $2\%$ at the $1\sigma$ level.
If $\delta \kappa_{\gamma}$ is measured to be larger than 4\%, the upper bound is obtained as $m_{\tilde\tau_1} < 200\GeV$.\footnote
{
The vacuum meta-stability condition determines the upper bound. If thermal transitions are taken into account, the constraint could be more severe especially when the stau is light \cite{Endo:2011uw}.
}  
Such a stau can be discovered at $500\GeV$ ILC. 
In fact, the stau is detectable up to $230\GeV$ at ILC with $\sqrt{s}=500\GeV$ and $\mathcal{L} = 500\invfb$ \cite{Baer:2013vqa}.
On the other hand, if $\delta \kappa_{\gamma}$ is measured to be $2\%$ ($1\%$), the stau mass is predicted to be less than $290\GeV$ ($460\GeV$). 
This is within the kinematical reach of $1\TeV$ ILC.
Therefore, if the stau contribution to $\kappa_{\gamma}$ is large enough to be measurable, the stau is predicted to be discovered at ILC.\footnote
{
Although the stau mass region could also be accessed by LHC, future sensitives of the stau searches have not been known. 
In particular, ILC is superior when the stau mass is degenerate with the Bino mass. 
} 

The above mass bounds weakly depend on $\tan\beta$. In the analysis, $\tan\beta = 20$ is chosen. 
If $\tan\beta$ increases, the lightest stau can be heavier, because the upper bound on $m_{\tilde{\tau}{LR}}^2$ from Eq.~\eqref{eq:VacuumStability} is relaxed. 
According to Ref.~\cite{Kitahara:2013lfa}, $\eta$ in Eq.~\eqref{eq:VacuumStability} increases as $\tan\beta$ becomes larger. For $\tan\beta = 70$, $\eta$ becomes unity, and the lightest stau is limited to be less than $220\GeV$ ($340\GeV$) for $\delta \kappa_{\gamma} = 4\%$ ($2\%$). Thus, the above conclusion does not change. 

Let us mention the case when the the heaviest stau is very heavy. 
In contrast to $\tilde\tau_1$, $\tilde\tau_2$ can be decoupled with $\kappa_{\gamma}$ enhanced and the vacuum meta-stability condition satisfied.
In Fig.~\ref{fig:StauMass}, $\delta \kappa_{\gamma}$ is insensitive to $m_{\tilde\tau_2}$ and determined by $m_{\tilde\tau_1}$ for very large $m_{\tilde\tau_2}$.
In the limit, $\mathcal{M}_{\gamma\gamma}(\tilde\tau)$ is determined only by $m_{\tilde\tau_1}$ and $g_{h\tilde\tau_{1}\tilde\tau_{1}}$. 
The vacuum meta-stability condition of $g_{h\tilde\tau_{1}\tilde\tau_{1}}$ is independent of $m_{\tilde\tau_2}$ and approximately proportional to $m_{\tilde\tau_1}$ \cite{Endo:2011uw}. 
Since the loop function $A_0^h (x_{\tilde\tau_1})$ is insensitive to $m_{\tilde\tau_1}$ for $m_{\tilde\tau_1}\gtrsim 100\GeV$, $\mathcal{M}_{\gamma\gamma}(\tilde\tau)$ is almost scaled by $1/m_{\tilde\tau_1}$, when the heaviest stau is decoupled. 
Thus, the excess of $\kappa_{\gamma}$ is explained by a light stau. As found in Fig.~\ref{fig:StauMass}, the upper bound on $m_{\tilde\tau_1}$ for larger $m_{\tilde\tau_2}$ is more severe than that for smaller $m_{\tilde\tau_2}$. Such a light stau can be discovered at ILC.

\section{Prospects of Stau} \label{sec:prospect}

Once the stau is discovered at ILC, its properties including the mass are determined. 
Especially, it is important to measure the stau mixing angle $\theta_{\tilde\tau}$. 
When $\sin 2\theta_{\tilde\tau}$ is sizable, the angle can be measured at ILC \cite{Nojiri:1996fp,Bechtle:2009em,Boos:2003vf,Endo:2013xka}. 
As observed in Fig.~\ref{fig:StauMass}, it is likely to be sizable to enhance $\kappa_{\gamma}$. 
In particular, if $\sin 2\theta_{\tilde\tau}$ is large enough to be measurable, the heaviest stau is likely to be light. 
Thus, it may be possible to discover the heaviest stau and measure its mass at ILC.
Then, the stau contribution to $\kappa_{\gamma}$ can be reconstructed by using the measured masses and mixing angle. 
This is a direct test whether the contribution is the origin of the deviation of $\kappa_{\gamma}$. 
On the other hand, the heaviest stau is not always discovered at the early stage of ILC, even if the stau mixing angle is measured. 
If $\theta_{\tilde\tau}$ as well as $m_{\tilde\tau_1}$ is measured, $m_{\tilde\tau_2}$ may be estimated in order to explain the excess of $\kappa_{\gamma}$.
In this section, we will study the reconstruction of the stau contribution to $\kappa_{\gamma}$. 
The mass of the heaviest stau and theoretical uncertainties will also be discussed.

\subsection{Reconstruction} \label{sec:reconst}

If both of $\tilde\tau_{1}$ and $\tilde\tau_{2}$ are measured, the stau contribution to $\kappa_{\gamma}$ can be reconstructed. The contribution is determined by the parameters in Eq.~\eqref{eq:parameters}. In this subsection, we discuss how and how accurately they are measured at ILC, and consequently the stau contribution to $\kappa_{\gamma}$ is reconstructed.

\begin{table}[t]
\centering
    \caption{Model parameters at our sample point. In addition, $\tan\beta = 5$ and $A_\tau = 0$ are set, though the results are almost independent of them.}
    \label{table:ModelPoint}
\vspace{0.5em}
    \begin{tabular}{l|cccc|c}
      \hline\hline
      {Parameters} & 
      $m_{\tilde{\tau}_1}$ & 
      $m_{\tilde{\tau}_2}$ & 
      $\sin \theta_{\tilde{\tau}}$& 
      $m_{\tilde{\chi}^0_1}$ & 
      $\delta \kappa_{\gamma}$\\
      \hline
      {Values} & 
      100\GeV & 230\GeV & 0.83 & 90\GeV & 
      3.6\% \\
      \hline\hline
    \end{tabular}
\end{table}

Let us first specify a model point to quantitatively study the accuracies. 
In table.~\ref{table:ModelPoint}, the stau masses, the stau mixing angle, and the Bino mass are shown. 
The point is not so far away from the SPS1a' benchmark point \cite{AguilarSaavedra:2005pw}, where ILC measurements have been studied (see e.g., Ref.~\cite{Bechtle:2009em}). 
The stau mixing angle is chosen to enhance the Higgs coupling as $\delta \kappa_{\gamma} = 3.6\%$. 
The staus masses are within the kinematical reach of ILC at $\sqrt{s}=500\GeV$. 
The point is consistent with the vacuum meta-stability condition and the current bounds from LHC and LEP. The most tight bound on the stau mass has been obtained at LEP as $m_{\tilde{\tau}_1} > 81.9\GeV$ at 95\% CL \cite{PDG}. LHC constraints are still weak \cite{ATLAS2013028}.
The other SUSY particles are simply supposed to be heavy. 
In particular, $\tan\beta = 5$ and $A_\tau = 0$ are chosen, where the Higgsino masses are about 2.2\TeV.

In order to reconstruct the stau contribution to $\kappa_{\gamma}$, it is required to measure the stau masses and the mixing angle.
At ILC, staus are produced in $e^+e^-$ collisions and decay into the tau and the Bino.
The stau masses are measured by studying the endpoints of the tau jets. 
In Ref.~\cite{Bechtle:2009em}, the mass measurement has been studied in detail at SPS1a'. 
It is argued that the mass can be measured at the accuracy of about $0.1\GeV$ ($6\GeV$) for $\tilde\tau_1$ ($\tilde\tau_2$).
Here, $\sqrt{s}=500\GeV$ and ${\cal L} = 500\invfb$ are assumed for ILC. 
The mass resolution may be improved by scanning the threshold productions \cite{Grannis:2002xd,Baer:2013cma}.
The accuracy could be $\sim 1\GeV$ for $m_{\tilde{\tau}_2} = 206\GeV$.
Since the model parameters of our sample point are not identical to those of SPS1a', the mass resolutions may be different from those estimated at SPS1a'. 
For instance, the production cross section of staus becomes different, while the SUSY background is negligible in our sample point. 
Profile of the tau jets depends on the masses of the staus and the Bino. 
In this letter, instead of analyzing the Monte Carlo simulation, we simply adopt the mass resolution,\footnote
{
The resolutions estimated in Ref.~\cite{Bechtle:2009em} depend on the uncertainty of the measured Bino mass. The Bino mass can be measured very precisely at ILC by the productions of selectrons or smuons \cite{Baer:2013vqa}, though they are irrelevant for $\kappa_{\gamma}$ and the vacuum meta-stability condition.
}
\begin{equation}
\Delta m_{\tilde{\tau}_1} \sim 0.1\GeV,~
\Delta m_{\tilde{\tau}_2} \sim 6\GeV.
\label{eq:ErrorMass}
\end{equation}

\begin{figure}[tb]
\begin{center}
 \includegraphics[width=7cm]{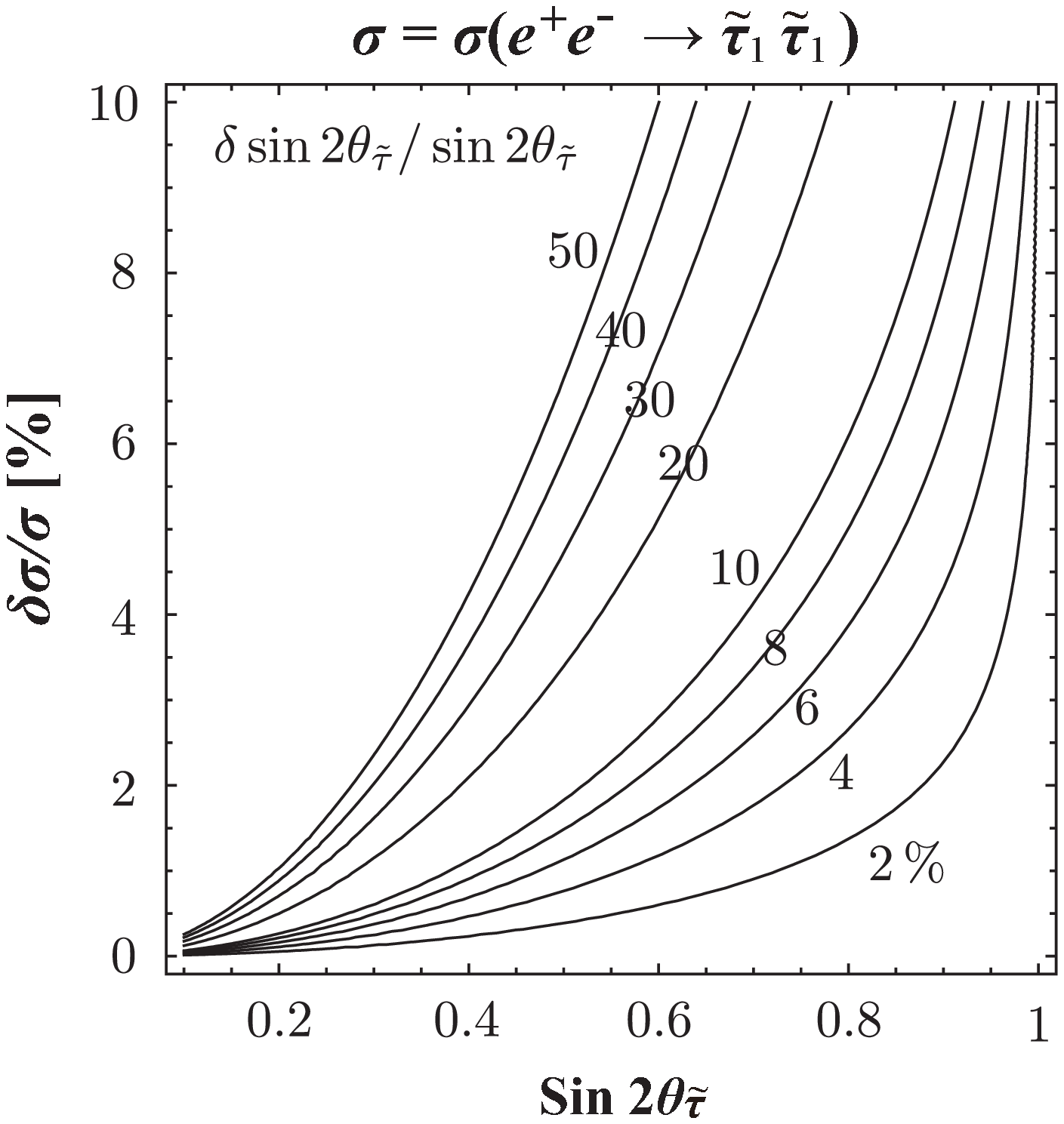} \hspace*{5mm}
 \includegraphics[width=7cm]{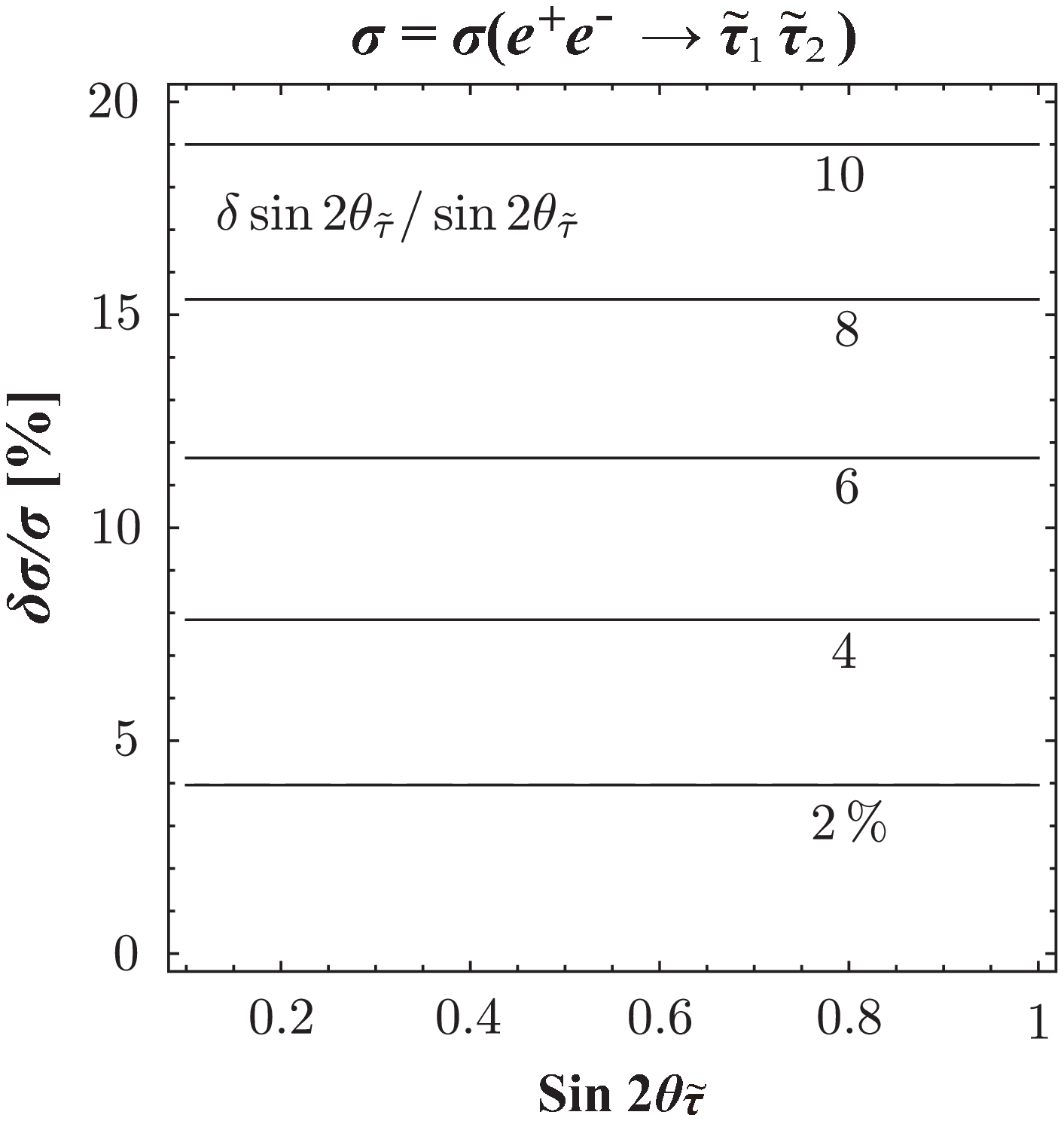}
 \end{center}
 \caption{Contours of $\delta \sin 2\theta_{\tilde\tau}/\sin 2\theta_{\tilde\tau}$ determined by the measurement of the production cross section of a pair of $\tilde\tau_1$ (\textbf{left}) and that of $\tilde\tau_1$ and $\tilde\tau_2$ (\textbf{right}). 
 Uncertainties from the mass resolutions are not taken into account.}
 \label{fig:StauMixingAngle}
\end{figure}

Next, let us discuss the measurement of the stau mixing angle, $\theta_{\tilde{\tau}}$.
Several methods have been studied for ILC. 
For instance, the polarization of the tau which is generated at the stau decay has been studied in Ref.~\cite{Nojiri:1996fp,Boos:2003vf,Bechtle:2009em}.
The angle can also be extracted from the production cross section of a pair of the lightest stau \cite{Boos:2003vf}. 
Note that accuracies of these angle measurement depend on the model point, i.e., the input value of $\theta_{\tilde\tau}$. 

In order to study the accuracy of the stau mixing angle at our sample point, let us investigate the production cross section of the lightest stau by following the procedure in Ref.~\cite{Endo:2013xka}.
The production cross section is given by \cite{Boos:2003vf}
\begin{align}
\sigma(e^+e^-\to\tilde\tau_1\tilde\tau_1) &= 
\frac{8\pi\alpha^2}{3s} \lambda^{\frac{3}{2}} 
\bigg[
  c_{11}^2 \frac{\Delta_Z^2}{\sin^42\theta_W}(P_{-+}L^2 + P_{+-}R^2)
\nonumber\\&~~~
  + \frac{1}{16} (P_{-+} + P_{+-})
  + c_{11}\frac{\Delta_Z}{2\sin^22\theta_W}(P_{-+}L + P_{+-}R)
\bigg],
\label{eq:CrossSection}
\end{align}
at the tree level, where the parameters are $\lambda = 1-4 m_{\tilde\tau_1}^2/s$, $\Delta_Z = s/(s-m_Z^2)$, and $c_{11} = [(L+R) + (L-R) \cos 2\theta_{\tilde{\tau}}]/2$ with $L=-1/2+\sin^2\theta_W$ and $R=\sin^2\theta_W$. 
The beam polarizations are parameterized as $P_{\mp\pm} = (1 \mp P_{e-}) (1 \pm P_{e+})$.
In the bracket, the first and second terms come from the s-channel exchange of the Z boson and the photon, respectively. The last term is induced by the interference of them.
The dependence on the stau mixing angle originates in the Z boson contribution. 

Since Eq.~\eqref{eq:CrossSection} is a function of the stau mass and mixing angle, $\theta_{\tilde\tau}$ is determined by measuring the cross section and the stau mass. 
In Fig.~\ref{fig:StauMixingAngle}, contours of the uncertainty of the stau mixing angle, $\delta \sin 2\theta_{\tilde\tau}/\sin 2\theta_{\tilde\tau}$, are shown.
In the left panel, the angle is determined from the production cross section of the lightest stau. 
The accuracy is sensitive to the input value of $\sin 2\theta_{\tilde\tau}$ and $\delta\sigma(\tilde\tau_1)/\sigma(\tilde\tau_1)$, where $\sigma(\tilde\tau_1) = \sigma(e^+e^-\to\tilde\tau_1\tilde\tau_1)$. 
In contrast, the uncertainty from the mass resolution of $\tilde\tau_1$ in Eq.~\eqref{eq:ErrorMass} is negligible.
The accuracy of $\sin 2\theta_{\tilde\tau}$ becomes better for larger $\sin 2\theta_{\tilde\tau}$.
If the stau contributes to $\kappa_{\gamma}$ sizably, $\sin 2\theta_{\tilde\tau}$ is likely to be large, as observed in Fig.~\ref{fig:StauMass}. 
Thus, the mixing angle is expected to be measured well.
At the sample point, where $\sin 2\theta_{\tilde\tau} = 0.92$, $\delta \sin 2\theta_{\tilde\tau}/\sin 2\theta_{\tilde\tau}$ is estimated to be better than 10\%, if the cross section is measured as precisely as $\delta\sigma(\tilde\tau_1)/\sigma(\tilde\tau_1) < 10\%$. 
At ILC, it is argued that the production cross section can be measured at the accuracy of about 3\%, according to the analysis in Ref.~\cite{Bechtle:2009em} at SPS1a'.
If $\delta\sigma(\tilde\tau_1)/\sigma(\tilde\tau_1) \sim 3\%$ is applied to our sample point, the accuracy is estimated to be
\begin{equation}
\Delta \sin 2\theta_{\tilde\tau}/\sin 2\theta_{\tilde\tau} \sim 2\%.
\label{eq:ErrorAngle}
\end{equation}

From Eqs.~\eqref{eq:ErrorMass} and \eqref{eq:ErrorAngle}, the accuracy of the reconstruction of the stau contribution to $\kappa_{\gamma}$ is estimated. 
If the errors are summed in quadrature, the uncertainty is obtained as
\begin{equation}
\Delta\kappa_{\gamma} \sim 0.5\%,
\label{eq:ErrorKappa}
\end{equation}
at the sample point, where $\delta \kappa_{\gamma} = 3.6\%$. 
Note that the uncertainty of the measurement of $\kappa_{\gamma}$ is 1--2\% from HL-LHC and ILC, as mentioned above.
Since the reconstruction error is comparable to or smaller than that of the measured $\kappa_{\gamma}$, it is possible to check whether the stau is the origin of the excess of the Higgs coupling $\kappa_{\gamma}$. 
It is emphasized that this is a direct test of the stau contribution to $\kappa_{\gamma}$. 

In Eq.~\eqref{eq:ErrorKappa}, the error is dominated by the uncertainties of the heaviest stau mass and the stau mixing angle. 
The former may be reduced by scanning the threshold of the stau productions, as mentioned above. 
For instance, if we adopt $\Delta m_{\tilde{\tau}_2} \sim 1\GeV$ as implied in Ref.~\cite{Grannis:2002xd,Baer:2013cma}, the error becomes $\Delta\kappa_{\gamma} \sim 0.3\%$. 
On the other hand, the latter uncertainty may be improved by studying the production cross section of $\tilde\tau_1$ and $\tilde\tau_2$ \cite{Bechtle:2009em}.
Since $e^+e^-\to\tilde\tau_1\tilde\tau_2$ proceeds by the s-channel exchange of the Z-boson, its cross section is proportional to $\sin^2 2\theta_{\tilde\tau}$ (see Ref.~\cite{Boos:2003vf} for the cross section).
Thus, it is very sensitive to the stau mixing angle, and further, the accuracy is independent of the model point, once the error of the production cross section is given. 
In the right panel of Fig.~\ref{fig:StauMixingAngle}, contours of $\delta \sin 2\theta_{\tilde\tau}/\sin 2\theta_{\tilde\tau}$ that is extracted from $\sigma(e^+e^-\to\tilde\tau_1\tilde\tau_2)$ are shown.
It is found that the accuracy is independent of the input $\sin 2\theta_{\tilde\tau}$. 
Here, uncertainties from the mass resolution are neglected. 
In particular, if the mass resolution of $\tilde\tau_2$ is large, the accuracy of the mixing angle becomes degraded. 
Unfortunately, the accuracy of the measurement of $\sigma(e^+e^-\to\tilde\tau_1\tilde\tau_2)$ has not been analyzed for ILC. 
Since $\sin 2\theta_{\tilde\tau}$ is likely to be large to enhance $\kappa_{\gamma}$, the cross section can be sizable.
At the sample point, it is estimated to be about 6\,\text{fb} for $\sqrt{s}=500\GeV$ with $(P_{e-},P_{e+})=(-0.8,0.3)$.
It is necessary to study this production process in future.

Let us comment on the $\tan\beta$ dependence.
At the sample point, $\tan\beta = 5$ is chosen. 
Although the stau contribution to $\kappa_{\gamma}$ includes $\tan\beta$, once the stau masses and mixing angle are measured, the reconstruction of the Higgs coupling is almost insensitive to it. 
This is because the stau left-right mixing parameter $m_{\tilde\tau{LR}}^2$ is determined by the measured masses and mixing angle through Eq.~\eqref{eq:MixingAngle}.
It can be checked that, even if $\tan\beta$ is varied, the accuracy \eqref{eq:ErrorKappa} is almost unchanged.

\subsection{Discussions} \label{sec:discussion}

Let us discuss miscellaneous prospects of the staus and theoretical uncertainties which have not been mentioned so far.

First of all, let us consider the situation when the heaviest stau is not discovered at the early stage of ILC, i.e., at $\sqrt{s}=500\GeV$. 
As found in Sec.~\ref{sec:mass}, if the excess of $\kappa_{\gamma}$ is measured at this stage, the lightest stau is already discovered. 
Then, it is possible to determine the stau mixing angle by measuring the production cross section of the lightest stau, as long as $\sin 2\theta_{\tilde\tau}$ is sizable (see Fig.~\ref{fig:StauMixingAngle}).
From the measurements of $m_{\tilde{\tau}_1}$, $\theta_{\tilde{\tau}}$ and $\kappa_{\gamma}$, the mass of the heaviest stau $m_{\tilde{\tau}_2}$ is determined. 
The predicted mass could be tested at the next stage of ILC, e.g., $\sqrt{s}=1\TeV$. 

In order to demonstrate the procedure, let us consider a model point with $m_{\tilde{\tau}_1}=150\GeV$, $m_{\tilde{\tau}_2}=400\GeV$ and $\sin\theta_{\tilde\tau}=0.54$. 
At the point, the Higgs coupling is $\delta \kappa_{\gamma}=5.6\%$.
At the early stage of ILC, it is expected that $\kappa_{\gamma}$ is determined with the uncertainty $\Delta\kappa_{\gamma} \sim 2\%$, and the lightest stau is measured with $\Delta m_{\tilde{\tau}_1} \sim 0.1\GeV$ and $\delta\sigma(\tilde\tau_1)/\sigma(\tilde\tau_1) \sim 3\%$. 
The stau mixing angle is extracted from the cross section as $\Delta \sin 2\theta_{\tilde\tau}/\sin 2\theta_{\tilde\tau} \sim 2.5\%$ (see Fig.~\ref{fig:StauMixingAngle}).
Since $\delta \kappa_{\gamma}$ is a function of $m_{\tilde{\tau}_1}$, $m_{\tilde{\tau}_2}$ and $\theta _{\tilde{\tau}}$, the mass of the heaviest stau is determined with the accuracy $\Delta m_{\tilde{\tau}_2} \sim 53\GeV$.
Here, the largest uncertainty comes from the measurement of the Higgs coupling.
If the error is reduced to be $\Delta\kappa_{\gamma} \sim 1\%$ due to reductions of the HL-LHC systematic uncertainties (see Ref.~\cite{Peskin:2013xra}), $\Delta m_{\tilde{\tau}_2} \sim 26\GeV$ is achieved. 
Such a prediction can be checked at ILC with $\sqrt{s}=1\TeV$.
This result would be helpful for choosing the beam energy to search for the heaviest stau at ILC.
Once $\tilde{\tau}_2$ is discovered, the stau contribution to the Higgs coupling can be reconstructed as Sec.~\ref{sec:reconst}. 

The uncertainty of the prediction of the heaviest stau mass depends on the model point, especially the stau mixing angle. 
If $m_{\tilde{\tau}_2}$ is larger, $\sin 2\theta_{\tilde\tau}$ is likely to be smaller, as expected from Fig.~\ref{fig:StauMass}. The measurement of the stau mixing angle, then, suffers from a larger uncertainty, and it becomes difficult to determine the mass of the heaviest stau. 

Next, let us mention extra contributions to the Higgs coupling from other SUSY particles. 
So far, they are suppressed because those particles are supposed to be heavy.
However, if the chargino, the stop or the sbottom is light, its contribution can be sizable \cite{Carena:2011aa,Batell:2013bka}. 
These particles are searched for effectively at (HL-) LHC (see e.g., Ref.~\cite{ATLAS:2013hta}).\footnote
{
On the other hand, it is possible to determine $\tan\beta$ by studying decays of staus, neutralinos or charginos at ILC, if the Higgsinos are light \cite{Boos:2003vf,Baer:2013cma}.
}
If none of them is discovered, their masses are bounded from below, and upper limits on their contributions to $\kappa_{\gamma}$ are derived.\footnote
{
If extra SUSY particles such as charginos are discovered, their contributions to $\kappa_{\gamma}$ may be reconstructed.
}
These extra contributions should be taken into account as a theoretical (systematic) uncertainty in the analysis of $\delta \kappa_{\gamma}$.

\begin{figure}[tb]
\begin{center}
 \includegraphics[width=7cm]{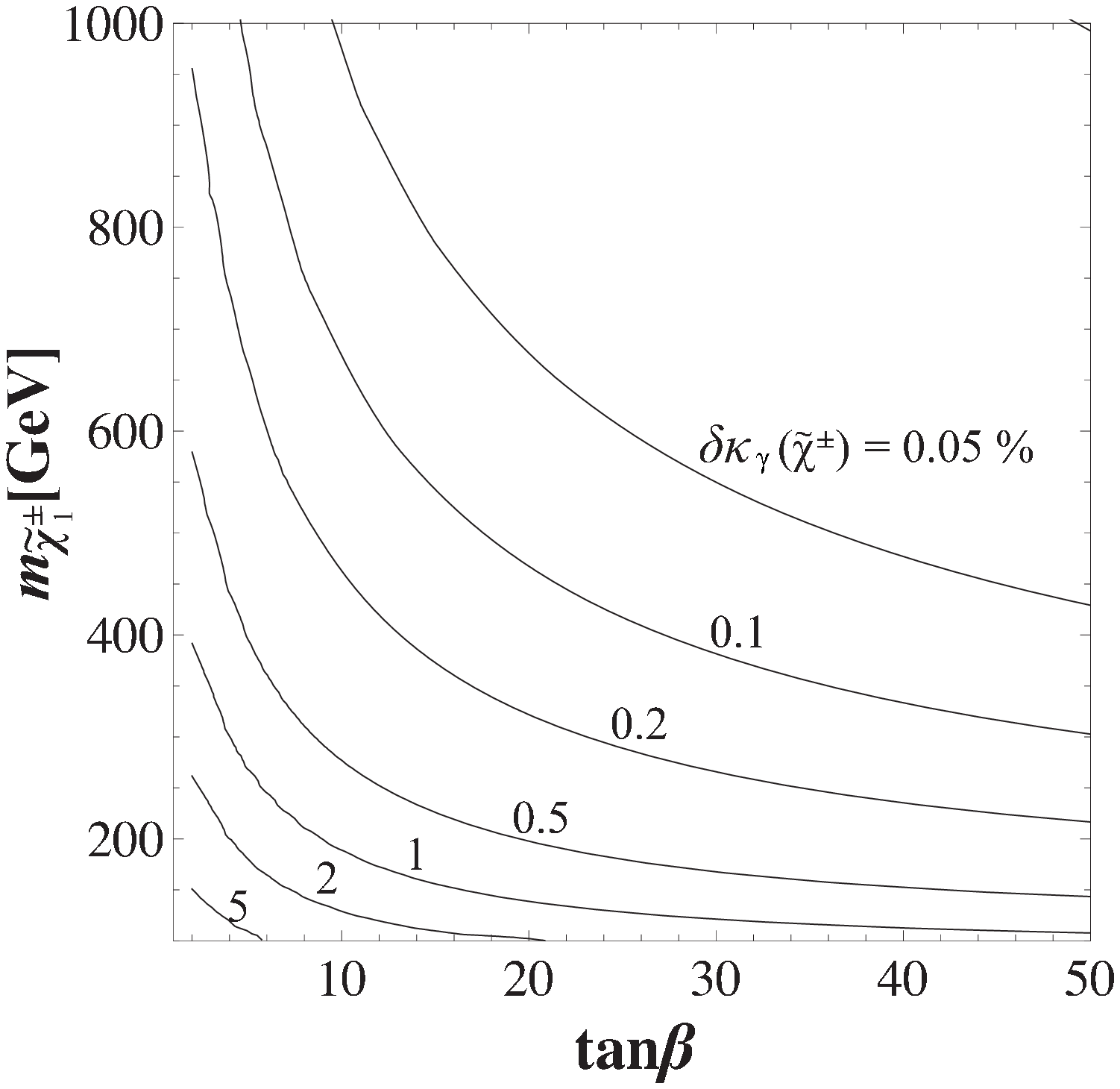} \hspace*{5mm}
 \includegraphics[width=7cm]{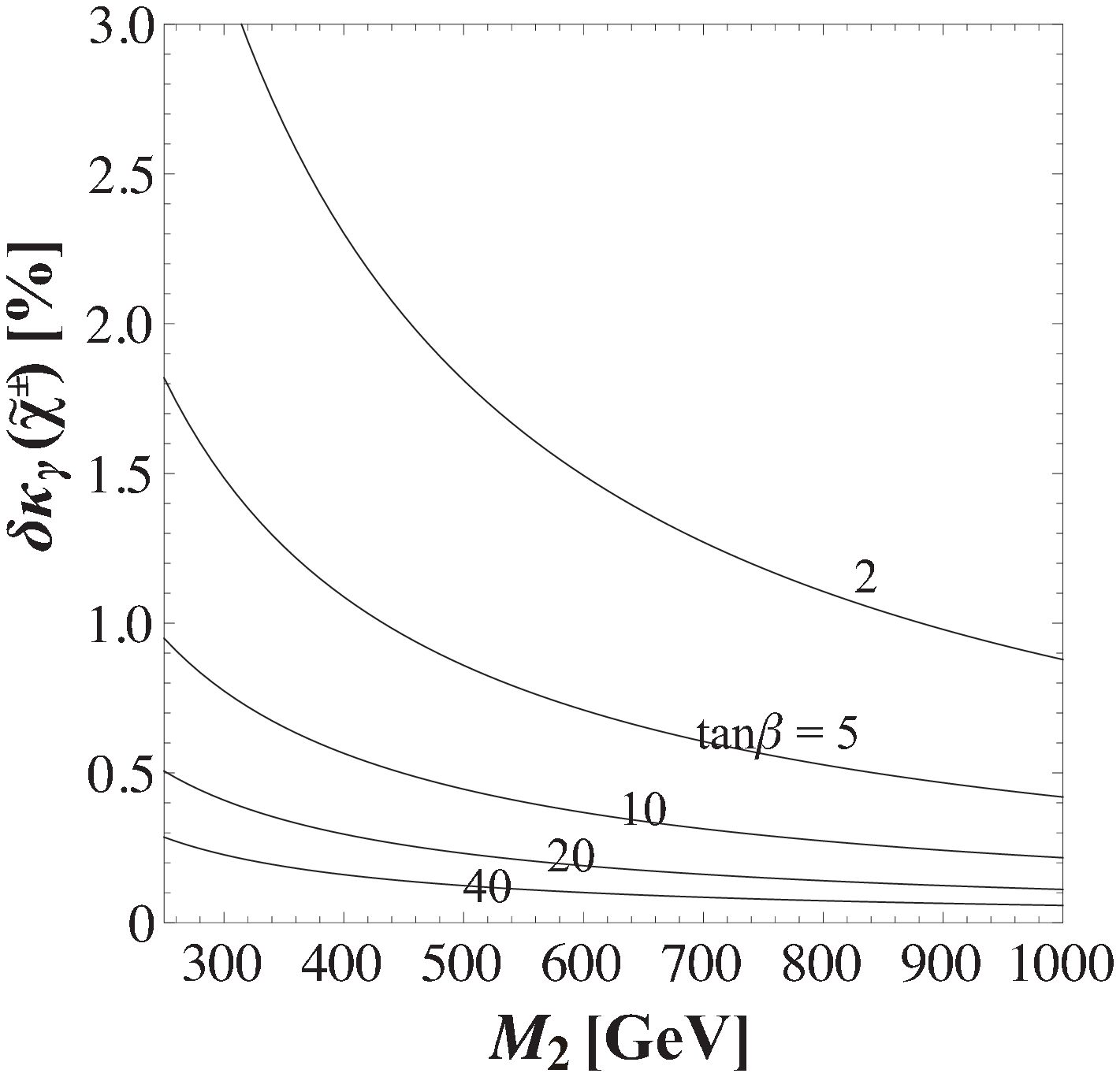}
 \end{center}
 \caption{Contours of the chargino contribution to the Higgs coupling $\delta \kappa_{\gamma}(\tilde\chi^\pm)$ are shown (\textbf{left}), where the Wino mass $M_2$ is equal to the Higgsino mass $\mu_H$. 
 The chargino contribution $\delta \kappa_{\gamma}(\tilde\chi^\pm)$ is displayed for various $\tan\beta$ (\textbf{right}), where $\mu_H$ is fixed to be $250\GeV$.
}
 \label{fig:Chargino}
\end{figure}

In order to estimate the theoretical uncertainty, extra contributions to $\kappa_{\gamma}$ are evaluated. 
In particular, since the experimental bounds on the chargino mass are still weak, the chargino contribution can be as large as the theoretical uncertainty.
At the one-loop level, the chargino contribution is given by \cite{Shifman:1979eb}
\begin{align} 
\mathcal{M}_{\gamma\gamma}(\tilde\chi^{\pm}) =
\sum_{i=1,2} \frac{ 2
g_{h\tilde\chi^{\pm}_i\tilde\chi^{\mp}_i}}{m_{\tilde\chi^{\pm}_i}}
A_{1/2}^h (x_{\tilde\chi^{\pm}_i}).
\end{align}
When the charginos are heavy, it is approximated as (c.f., Ref.~\cite{Batell:2013bka})
\begin{align}
\mathcal{M}_{\gamma\gamma}(\tilde\chi^{\pm}) = 
\frac{4}{3}
\frac{g^2 v \sin 2 \beta}{M_2 \mu_{H} - \frac{1}{4} g^2 v^2 \sin 2 \beta}.
\label{eq:MgammaChargino}
\end{align}
In the left panel of Fig.~\ref{fig:Chargino}, contours of $\delta \kappa_{\gamma}(\tilde\chi^\pm)$ which is induced by the charginos are displayed. 
Here we take  $M_2 = \mu_H$. 
The contribution decreases as $m_{\tilde\chi^\pm_1}$ or $\tan\beta$ increases.
If the charginos are constrained to be heavier than $600\GeV$ ($1\TeV$), these contribution to $\kappa_{\gamma}$ is estimated to be smaller than $0.5\%$ ($0.2\%$) for $\tan\beta > 2$.\footnote{The chargino contribution is unlikely to dominate the contributions to $\delta \kappa_{\gamma}$, unless it is very light and $\tan\beta$ is small.}
This is considered to be a theoretical uncertainty. 
In addition, if either of the Wino or the Higgsino is decoupled, $\delta \kappa_{\gamma}(\tilde\chi^\pm)$ is suppressed.
Such a feature is observed in the right panel of Fig.~\ref{fig:Chargino}, where $\mu_H$ is fixed to be $250\GeV$ for various $\tan\beta$.

The stop and the sbottom can also contribute to $\kappa_{\gamma}$ sizably \cite{Carena:2011aa}. 
It should be noted that they simultaneously modify the Higgs coupling to di-gluon. 
The coupling $\kappa_{g}$ is measured precisely at ILC at the (sub) percent levels \cite{Asner:2013psa,Peskin:2013xra}.
Thus, if deviations are discovered in $\kappa_{g}$ as well as $\kappa_{\gamma}$, it is interesting to study the contributions of the stop or the sbottom.

\section{Conclusion} \label{sec:conclusion}

In this letter, the stau contribution to the Higgs coupling to di-photon was studied. 
The coupling $\kappa_{\gamma}$ will be measured at the percent levels by the joint analysis of HL-LHC and ILC. 
Such precise measurements may enable us to detect effects of the new charged particles that couple to the Higgs boson such as the stau. 
In this letter, we first studied the stau mass region by taking the vacuum meta-stability condition into account. 
Consequently, we found that, if the excess of $\kappa_{\gamma}$ is measured to be larger  than $4\%$ at the early stage of ILC ($\sqrt{s}=500\GeV$), the lightest stau is predicted to be lighter than about $200\GeV$. Such a stau can be discovered at ILC.
Also, it was shown that, if the excess of $\kappa_{\gamma}$ is measured to be $1\text{--}2\%$ by accumulating the luminosity at $1\TeV$ ILC, the lightest stau mass is bounded to be less than $290\text{--}460\GeV$. 
This stau is within the kinematical reach of ILC.
Therefore, we concluded that the stau contribution to $\kappa_{\gamma}$ can be probed by discovering the stau, if the excess of $\kappa_{\gamma}$ is measured in the future experiments, and if it originates in the stau contribution.

Once the stau is discovered at ILC, its properties are determined precisely.
In this letter, we also studied the reconstruction of the stau contribution to $\kappa_{\gamma}$ by using the information which is available at ILC. 
It was estimated that the contribution can be reconstructed at $\sim 0.5\%$ at the sample point, which is comparable to or smaller than the measured value of the Higgs coupling. 
Thus, it is possible to test directly whether the excess originates in the stau contribution.
Here, the measurement of the stau mixing angle is crucial. 
We also argued that, if the stau mixing angle is measured at the early stage of ILC, it is also possible to predict the heaviest stau mass, even when the heaviest stau is not yet discovered at the moment. 
Therefore, the stau contribution to $\kappa_{\gamma}$ can be probed not only by discovering the lightest stau, but also by studying the stau properties. 

Discoveries of new physics  are the next target after the discovery of the Higgs boson. 
The measurement of the Higgs couplings to di-photon is one of the hopeful channels to search for the new physics. 
The stau contribution to the Higgs coupling could be probed or tested in future colliders by following the analysis in this letter.

\section*{Acknowledgements}
This work was supported by JSPS KAKENHI Grant No.~23740172 (M.E.) and 25-10486 (T.Y.).
The work of T.Y. is supported in part by a JSPS Research Fellowship for Young Scientists, and
supported by an Advanced Leading Graduate Course for Photon Science grant.

\providecommand{\href}[2]{#2}
\begingroup\raggedright
\endgroup

\end{document}